# Current-induced magnetoresistance hysteresis in the kagome superconductor $CsV_3Sb_5$

Han-Xin Lou,[1] Xing-Guo Ye,[1] Xin Liao,[1] Qing Yin,[1] Da-Peng Yu,[2,3,4] and Zhi-Min Liao[1,2,*]

[1]*State Key Laboratory for Mesoscopic Physics and Frontiers Science Center for Nano-optoelectronics,*
*School of Physics, Peking University, Beijing 100871, China*
[2]*Hefei National Laboratory, Hefei 230088, China*
[3]*Shenzhen Institute for Quantum Science and Engineering, Southern University of Science and Technology, Shenzhen 518055, China*
[4]*International Quantum Academy, Shenzhen 518048, China*

We report the observation of current-modulated magnetoresistance hysteresis below the superconducting transition temperature in the kagome superconductor $CsV_3Sb_5$. This highly tunable hysteresis behavior is confined to the superconducting state and vanishes when superconductivity is fully suppressed, directly linking magnetoresistance hysteresis to the superconducting order in $CsV_3Sb_5$. Additionally, the superconducting diode effect driven by a small magnetic field is observed, indicating the enhanced electronic magnetochiral anisotropy by the chiral domain-wall scattering. Our findings position $CsV_3Sb_5$ as a promising platform for exploring nontrivial physical phenomena, including unconventional pairing mechanisms and topological superconductivity.

## I. INTRODUCTION

In recent years, the quasi-two-dimensional layered kagome metals $AV_3Sb_5$ ($A$ = K, Rb, Cs) have garnered significant attention due to their multiple exotic electronic states and rich phase diagram [1–13]. With the coexistence of various charge order states and multigap superconductivity [14–19], the $AV_3Sb_5$ family presents a different platform for exploring unconventional superconductivity and exotic pairing mechanisms [20–28]. Notably, time-reversal symmetry breaking has been observed in the superconducting states of these nonmagnetic materials [10,11,29–31], suggesting the presence of chiral superconductivity [21,22]. Theoretical studies have proposed several mechanisms for this time-reversal symmetry breaking, including chiral charge order [21,32,33] and loop current [34–36]. Experimentally, zero-field muon spin relaxation experiments in $KV_3Sb_5$ reveal the emergence of a local magnetic field below the charge density wave transition temperature, persisting into the superconducting state [29]. Additionally, a giant anomalous Hall effect observed in both $KV_3Sb_5$ and $CsV_3Sb_5$ below the charge density wave transition temperature further indicates time-reversal symmetry-breaking charge orders [37–39]. Recent studies have also reported the switchable chiral transport [32] and a zero-field superconducting diode effect in $CsV_3Sb_5$ [22], demonstrating the existence of dynamic superconducting domains. The interplay between time-reversal symmetry breaking and superconductivity in kagome materials opens potential avenues for topological quantum computations [40,41]. Despite significant advances in exploring time-reversal symmetry breaking in $CsV_3Sb_5$ [22,29,30–32,37–39,42], further research on modulating this symmetry breaking is crucial for a deeper understanding of the superconducting properties without time-reversal symmetry and could pave the way for future device applications.

In this work, we report current-induced magnetoresistance hysteresis below the superconducting transition temperature ($T_c$) in the kagome superconductor $CsV_3Sb_5$ through electrical transport measurement, highlighting the symbiotic relationship between magnetoresistance hysteresis and superconducting states. Upon applying a DC current ($I_{DC}$), the hysteresis loop is observed when sweeping the magnetic field forward and backward. Notably, this magnetoresistance hysteresis can be modulated by both $I_{DC}$ and temperature. Additionally, a pronounced superconducting diode effect is investigated under a small magnetic field applied along both out-of-plane and in-plane directions, showing nonreciprocity in superconducting current transmission, which can be significantly enhanced and highly tuned by the external magnetic fields. We discuss the possible origins of the magnetoresistance hysteresis and the behind mechanisms underlying these modulations, including the potential roles of loop current, vortex physics, and chiral superconducting domains in $CsV_3Sb_5$.

## II. EXPERIMENTAL RESULTS

The results from three devices, labeled S1–S3, are presented in this work. Figure 1 highlights the basic superconducting properties of device S1. As shown in Fig. 1(a), a thin $CsV_3Sb_5$ flake, less than 30 nm thick, was exfoliated from a bulk crystal using the mechanical method and placed onto pre-fabricated electrodes (see Appendix A). The resistance ($R$) vs temperature ($T$) curve of device S1, measured using the four-probe method with an AC current of $I_{AC} = 1$ μA, shows a superconducting transition beginning at ∼4.2 K [Fig. 1(b)], consistent with previous reports. Figure 1(c) illustrates the impact of an out-of-plane magnetic field on the superconducting state at 1.8 K, with an onset critical field $B_c \sim 0.6$ T. The superconducting transition under the magnetic field is gradual but consistent, with complete suppression at 1 T. Figure 1(d) presents the differential resistance ($dV/dI$) as a function of the applied DC current $I_{DC}$ at 1.8 K, with an onset critical current $I_c$ around ±60 μA. The curves obtained from $I_{DC}$ sweeps, from 400 to −400 μA (red line) and reverse (black line), show complete overlaps. The unsharp transition with multiple peaks suggests the possible existence of superconducting domains [22].

*Contact author: liaozm@pku.edu.cn

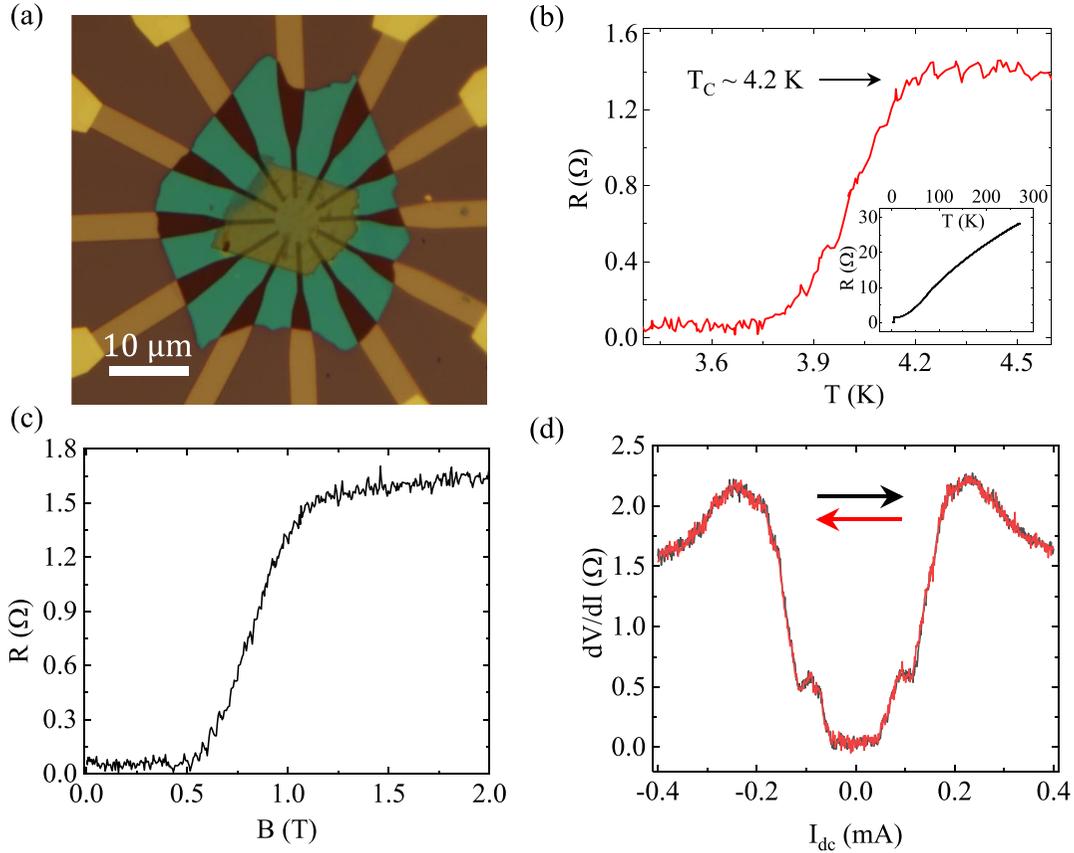

FIG. 1. Basic transport properties of CsV$_3$Sb$_5$ device S1. (a) Optical image of a typical CsV$_3$Sb$_5$ device, with a thin flake thickness of less than 30 nm. (b) Temperature-dependent resistance measured by four-probe method, showing $T_c \sim 4.2$ K. (c) Out-of-plane magnetic field dependence of resistance at 1.8 K, with an onset $B_c \sim 0.6$ T. (d) Differential resistance as a function of direct current at 1.8 K and $I_{AC} = 1$ μA. The black and red curves represent the current sweeping from −400 to 400 μA and the reverse, respectively.

As shown in Fig. 2(a), the magnetic field dependent differential resistance ($dV/dI$) at 1.8 K is almost identical for forward (black line) and backward (red line) sweeps when no direct current $I_{DC}$ is applied, with no visible magnetoresistance hysteresis. However, when a 40 μA direct current is applied, the magnetoresistance curves exhibit a clear hysteresis loop [Fig. 2(b)], with the critical field $B_c$ being larger for the sweep from high field to zero compared to the sweep from zero to high field. Figures 2(c) and 2(d) show the magnetoresistance hysteresis under unilateral magnetic field scans between 0 to 1 T and 0 to −1 T at $I_{DC} = 60$ μA, respectively. To further investigate the influence of $I_{DC}$ on magnetoresistance hysteresis, a range of currents 0–80 μA was applied to modulate the superconducting transition in CsV$_3$Sb$_5$. Figure 2(e) shows the magnetoresistance measurements under different direct currents, with the magnetic field swept from

−1 to 1 T and then back from 1 to −1 T in each set. For $I_{DC}$ below 10 μA, there is almost no observable hysteresis between the curves for opposite sweep directions. However, as $I_{DC}$ increases, the magnetoresistance hysteresis loops become more pronounced, accompanied by a decrease in the critical magnetic field. The difference in the critical field between forward and backward magnetic field sweeps is denoted as $\Delta B_c$, where the detailed definition is provided in Appendix B. The $\Delta B_c$ shows a monotonic increase with $I_{DC}$, as depicted in Fig. 2(f). These results obtained with a positive $I_{DC}$ are consistent with those using a negative $I_{DC}$, as depicted in Fig. 6 (see Appendix C).

We further investigated the magnetoresistance hysteresis behavior at different temperatures in device S1. As shown in Fig. 3(a), with $I_{DC} = 10$ μA, the hysteresis loop remains visible at 3 K. However, as the temperature increases and superconductivity is completely suppressed, the hysteresis disappears. This variation in magnetoresistance hysteresis aligns with the chiral superconducting domains in CsV$_3$Sb$_5$ [43,44]. To rule out the possibility of artificial hysteresis caused by data acquisition lag—where a faster field sweep rate could lead to apparent hysteresis—we conducted measurements at sweeping rates 15, 20, and 25 mT/min, as shown in Fig. 3(b). The three resulting curves overlap well and consistently display clear magnetoresistance hysteresis, indicating that this effect is not sensitive to sweeping rates. Therefore, the possibility of artificial hysteresis can be excluded. In addition, the thickness of device S2 is around 45 nm, where the magnetoresistance hysteresis loop remains visible but the difference of critical field $\Delta B_c$ is smaller than that of the thin sample S1. The observation in samples with various thickness indicates that the magnetoresistance hysteresis behavior may be insensitive within a certain thickness range, the magnitude of which varies with the sample becoming thicker [45].

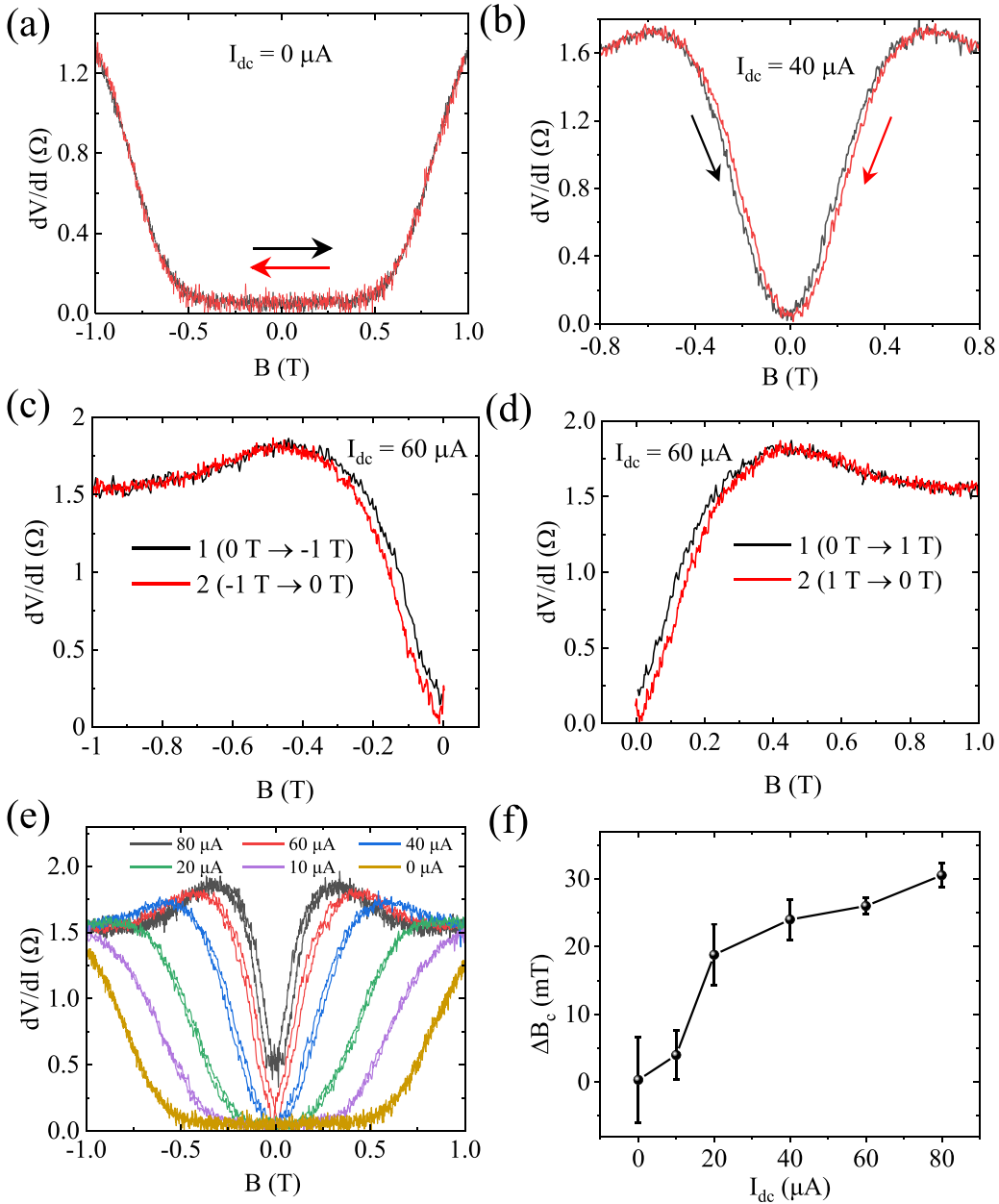

FIG. 2. Magnetoresistance hysteresis below superconducting transition temperatures in device S1 at $T = 1.8$ K and $I_{AC} = 1$ μA. (a) Out-of-plane magnetic field dependence of magnetoresistance. The black and red curves represent magnetic field sweeps from $-1$ to $1$ T and reverse, respectively. (b) Magnetoresistance hysteresis at $I_{DC} = 40$ μA. (c),(d) Magnetoresistance hysteresis at $I_{DC} = 60$ μA clearly display the behavior in negative and positive magnetic fields, respectively. (e) Magnetoresistance curves with varying $I_{DC}$. (f) Current dependence of the magnetic field difference $\Delta B_c$ with error bars obtained from (e), showing a monotonically increasing trend.

To gain deeper insight into the unconventional superconductivity in $CsV_3Sb_5$, we investigate the superconducting diode effect, which refers to the nonreciprocity of superconducting current transmission in opposite directions and is often associated with symmetry breaking. The zero-field superconducting diode effect in $CsV_3Sb_5$, recently reported in Ref. [22], suggests the presence of chiral superconductivity domains. In our measurements, this effect is also observed in device S3, though it is relatively weak (see Appendix D). Here, we focus on the modulation of nonreciprocal superconducting current by applying a small magnetic field, where the superconducting diode effect is significantly enhanced. Figure 4 shows the differential resistance in device S3 as a function of $I_{DC}$ at different magnetic fields. For ease of comparison, the measurement results for negative current sweeping are symmetrically plotted on the same side as the positive current. The black and red curves represent current sweeping from 0 to 200 μA and 0 to −200 μA, respectively. Due to the sensitivity of superconductivity to an out-of-plane magnetic field, a pronounced superconducting diode effect is observed at a small magnetic field of 5 mT [Fig. 4(a)]. The critical current $I_c^-$ in the negative sweeping direction is larger than $I_c^+$ in the positive direction with a positive magnetic field applied. When the magnetic field polarity is reversed, the sign of the superconducting diode effect also reverses, as shown in Fig. 4(b). As the magnetic field increases, the difference $\Delta I_c$ between $I_c^+$ and $I_c^-$ (where $\Delta I_c = I_c^+ - I_c^-$) becomes more pronounced, though the polarity of the superconducting diode effect remains unchanged. Figure 4(c) shows the dependence of $I_c$ on the out-of-plane magnetic field $B$. The response of $I_c$ to positive and negative magnetic fields is notably different: $I_c^+$ ($I_c^-$) decreases monotonically with increasing positive (negative) magnetic field, while remaining relatively stable on the opposite side.

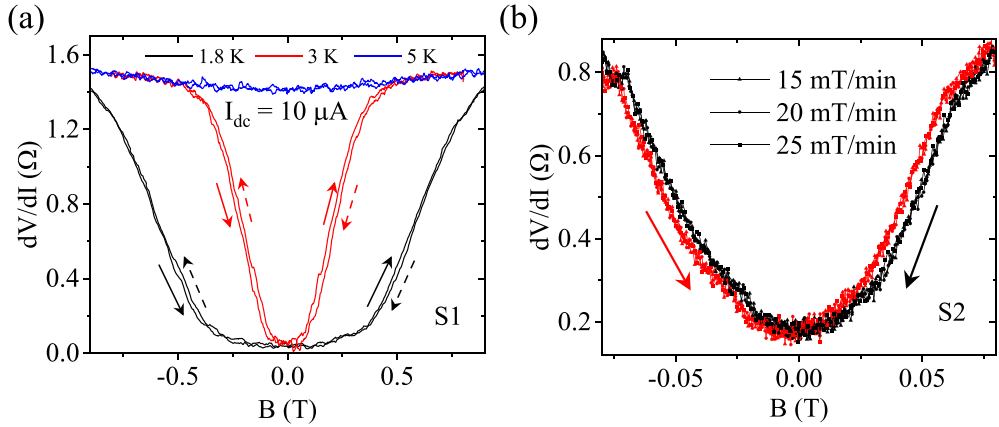

FIG. 3. Magnetoresistance under varying temperatures and field sweep rates. (a) Magnetoresistance hysteresis measured at different temperatures in device S1 with $I_{AC} = 1$ μA and $I_{DC} = 10$ μA. (b) Magnetoresistance hysteresis in device S2 at $I_{AC} = 2$ μA and $I_{DC} = 60$ μA under different field sweep rates.

The nonreciprocal superconducting current transmission under magnetic field has the origin of the electronic magnetochiral anisotropy (eMChA), which can lead to the asymmetric critical currents in opposite current directions [32]. The eMChA is conventionally expressed as $R(I, B) = R_0(1 + \mu^2 B^2 + \gamma IB)$, usually referring to the resistance change in response to the applied current $I$ and external magnetic field $B$, where $\mu$ is the mobility and $\gamma$ is the eMChA coefficient. Although the characteristics of the eMChA effect is allowed in many systems with space-reflection symmetry breaking, it is commonly too weak to reach an appreciable level. Here, the strongly enhanced eMChA response in the superconducting state aligns with the emergence of chiral domains, where the domain walls act as ideal chiral scattering centers and can be highly tunable and switchable at low magnetic field [32]. To further explore the influence of magnetic field orientation on superconducting current transmission in $CsV_3Sb_5$, Figs. 4(d)–4(f) display the differential resistance curves under various in-plane magnetic fields. The superconducting diode effect is clearly observed at a larger in-plane magnetic field of 100 mT, exhibiting the same polarity behavior as with the out-of-plane magnetic field. However, the response of the superconductivity to the out-of-plane magnetic field is much stronger than to the in-plane magnetic field.

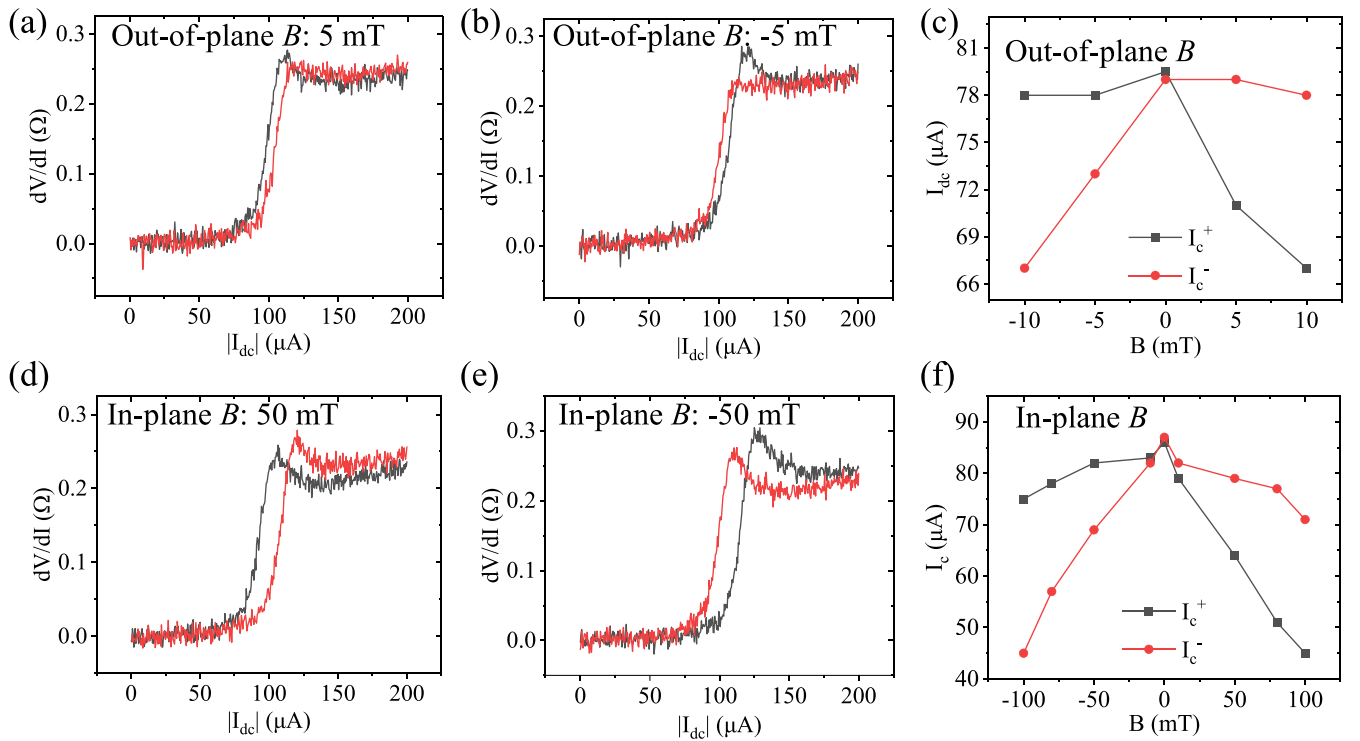

FIG. 4. Superconducting diode effect in device S3 at 1.5 K. Panels (a) and (b) display the $dV/dI$ as a function of $I_{DC}$ under an out-of-plane magnetic field. The black and red curves correspond to measurements taken with $I_{DC}$ varying from 0 to 200 μA and 0 to −200 μA, respectively, illustrating the superconducting diode effect. Panels (d) and (e) show $dV/dI$ as a function of $I_{DC}$ under an in-plane magnetic field. Panels (c) and (f) present the magnetic field dependence of the critical current $I_c$ for positive and negative current bias regimes.

## III. DISCUSSION

We next discuss the potential mechanisms underlying the observed magnetoresistance hysteresis in $CsV_3Sb_5$. We rule out the residual magnetization of the system magnet as the cause of these phenomena. This is because the residual magnetization does not vary with the sample temperature or applied current, whereas our results show that the magnetoresistance hysteresis is modulated by both. Additionally, while residual magnetization typically depends on the magnetic field sweep rate, our observed magnetoresistance hysteresis does not. More detailed discussions to exclude other extrinsic origins are provided in Appendix E.

Intrinsic mechanisms that could lead to magnetoresistance hysteresis during the superconducting transition include ferromagnetic orders, chiral superconducting domains, vortex physics, and spin-triplet pairing [43,46–50]. Since the magnetoresistance hysteresis is directly modulated by the applied current, it is unlikely to be caused by long-range ferromagnetic order [48]. And also, spectroscopy measurements in $CsV_3Sb_5$ provide no evidence of long-range magnetic order [1,51]. The vortex pinning effect is one of the potential reasons for the magnetoresistance hysteresis behavior, but the pinning strength in our devices is weak (see Appendix E) [52]. Together with the superconducting diode effect driven by the small magnetic fields, the magnetoresistance hysteresis in our study aligns more closely with the concept of chiral superconducting domains in $CsV_3Sb_5$. The loop currents with opposite chirality may become unbalanced due to strain introduced during sample preparation or material inhomogeneity, leading to the formation of chiral superconducting domains. Additionally, applying a magnetic field can align the chirality of the domains, leading to a discrepancy between magnetoresistance curves for opposite sweep directions, which manifests as hysteresis. When the applied current is increased without fully destroying the superconductivity, the density of loop currents is enhanced, thereby making the magnetoresistance hysteresis more pronounced. Moreover, the emergence of chiral domains can lead to the enhanced eMChA effect, resulting in the magnetic field-driven superconducting diode effect.

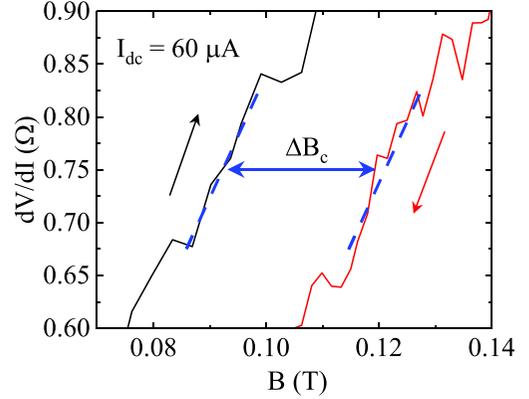

FIG. 5. The definition of the difference of magnetic field $\Delta B_c$. The magnetoresistance hysteresis loop measured at $T = 1.8$ K, $I_{AC} = 1$ μA and $I_{DC} = 60$ μA, where $\Delta B_c$ is the difference of magnetic fields swept from opposite directions. The black and red curves are measured by sweeping the magnetic field forward and backward, respectively.

## IV. CONCLUSION

In summary, we have observed magnetoresistance hysteresis and the superconducting diode effect in the superconducting state of $CsV_3Sb_5$ through transport measurements, providing experimental evidence for superconductivity with symmetry breaking. When a direct current is applied, the magnetoresistance curve swept from high field to zero shows a larger critical magnetic field than when swept in the opposite direction. Notably, the dependence of magnetoresistance hysteresis on direct current and temperature highlights the tunability of the superconducting transition in $CsV_3Sb_5$. Together with the magnetic field dependent nonreciprocity associated with the eMChA effect, the existence of chiral superconducting domains in the superconducting state provides a plausible explanation. These findings shed light on the potential origins and intrinsic modulation mechanisms of chiral superconductivity, offering interesting prospects for tuning kagome superconductors with time-reversal symmetry-breaking superconducting order.

## ACKNOWLEDGMENTS

This work was supported by the Innovation Program for Quantum Science and Technology (Grant No. 2021ZD0302403) and the National Natural Science Foundation of China (Grants No. 62425401 and No. 62321004).

## APPENDIX A: DEVICE FABRICATION AND MEASUREMENTS

*Device fabrication.* Employing the standard mechanically exfoliated method, we obtained $CsV_3Sb_5$ thin flakes from the bulk crystals. The Ti/Au electrodes (2/10 nm thick) were prefabricated on the $SiO_2$/Si substrate by electron beam evaporation, and the thin layer $CsV_3Sb_5$ flakes were transferred onto the electrodes sing a polymer-based dry transfer technique. The device is encapsulated by the flake of hexagonal boron nitride (hBN) to prevent oxidation of $CsV_3Sb_5$. The whole exfoliating and transfer process was done in an argon-filled glove box with $O_2$ and $H_2O$ content below 0.01 parts per million to avoid sample degeneration.

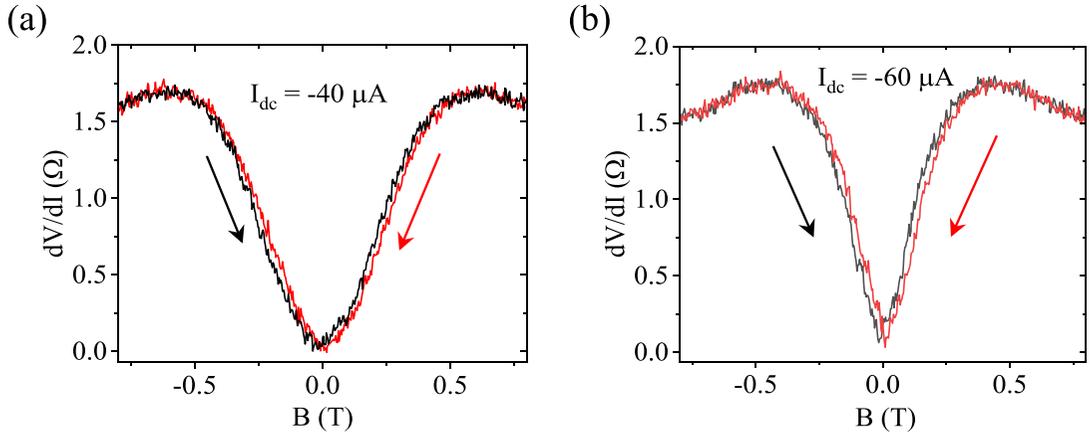

FIG. 6. Magnetoresistance hysteresis induced by negative DC current in device S1. (a),(b) The magnetoresistance hysteresis is shown at $T = 1.8$ K, (a) $I_{DC} = -40$ μA and (b) $I_{DC} = -60$ μA, respectively. The black and red curves represent the magnetic field sweeping from $-1$ to 1 T and reverse, respectively.

*Transport measurements.* Transport measurements were carried out in an Oxford cryostat with a variable temperature insert and a superconducting magnet. First-harmonic signals were collected by standard lock-in techniques (Stanford Research Systems Model SR830) with frequency $\omega$. Frequency $\omega$ equals 17.777 Hz unless otherwise stated. The direct current was applied through Keithley 2400 SourceMeters.

## APPENDIX B: DEFINITION OF THE DIFFERENCE OF MAGNETIC FIELD

As shown in Fig. 5, we define the $\Delta B_c = B_{c1} - B_{c2}$, where $B_{c1}$ and $B_{c2}$ are the average magnetic fields at which the $dV/dI$ becomes 45–55% of the normal resistance in the positive magnetic field regime for backward and forward sweeping, respectively.

## APPENDIX C: MAGNETORESISTANCE HYSTERESIS INDUCED BY NEGATIVE DC CURRENT

As shown in Fig. 6, the experimental results obtained by applying negative DC current are consistent with those using a positive $I_{DC}$, presenting magnetoresistance hysteresis behavior.

## APPENDIX D: ZERO MAGNETIC FIELD SUPERCONDUCTING DIODES

The zero magnetic field superconducting diode effect in $CsV_3Sb_5$ is observed in our measurements, though the signal is relatively weak (Fig. 7).

## APPENDIX E: OTHER POSSIBLE MECHANISMS FOR THE MAGNETORESISTANCE HYSTERESIS

There are several mechanisms for the magnetoresistance hysteresis. Here we made corresponding discussions and ruled them out.

### 1. Magmatic field sweep induced heating effect

Magnetoresistance hysteresis can be caused by the magnetic field sweep induced heating effect. Heating would be induced when sweeping the magnetic field. Importantly, the heating effect is stronger when decreasing the magnetic field from high values to zero, rather than the case when increasing the magnetic field from zero to high values. Therefore, artificial hysteresis can be induced through such heating effect. However, in the scenario of magnetic field sweep induced heating effect, the absolute value of critical field $B_c$ should be larger when sweeping field from zero increasing to high magnetic fields, rather than sweeping inversely, since the heating effect is stronger when increasing the magnetic field to high values. Comparing with the magnetoresistance hysteresis curves observed in our measurements (Fig. 2), the heating effect induced by magnetic field sweep can be safely excluded.

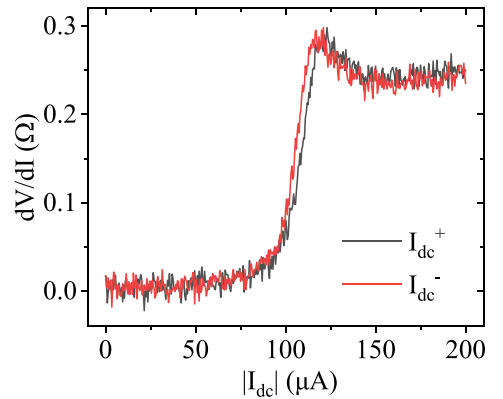

FIG. 7. Superconducting diode in device S3. $dV/dI$ as a function of $I_{DC}$ without applied magnetic field. The black curves and red curves are measured by $I_{DC}^+$ ($0 \to 200$ μA) and $I_{DC}^-$ ($0 \to -200$ μA), respectively, presenting the signal of superconducting diode effect.

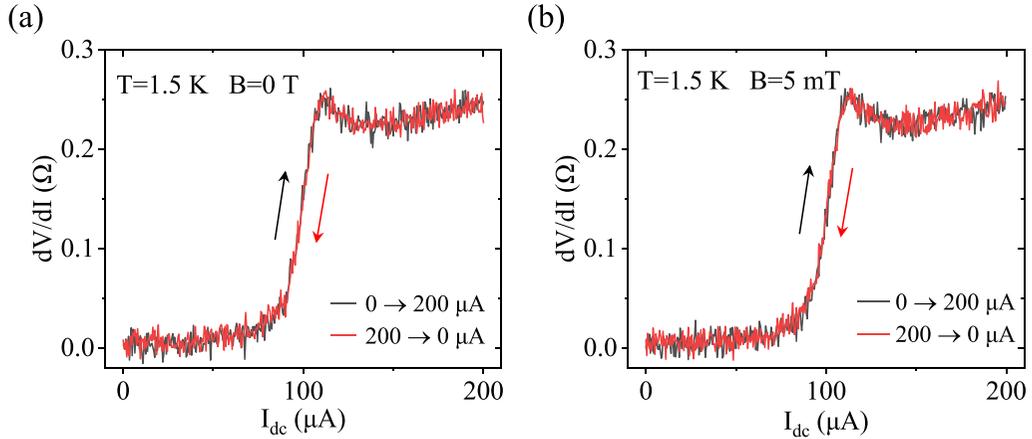

FIG. 8. *I-V* curves with opposite sweeping directions in S3. Differential resistance $dV/dI$ as a function of direct current at $T = 1.5$ K, with out-of-plane magnetic fields $B = 0$ T (a) and $B = 5$ mT (b). The black and red curves represent the current sweeping from 0 to 200 μA and reverse, respectively. The curves obtained from opposite $I_{DC}$ sweeping directions show complete overlaps.

### 2. Flux trapping

The flux trapping mechanism is another reason for the magnetoresistance hysteresis in a superconducting transition, the overall hysteresis loop caused by which will always show a butterflylike pattern [53,54]. In this scenario, because of the existence of weak flux pining sites, the magnetoresistance will present a significant peak at a very small magnetic field when sweeping from zero to high fields. In addition, in the process of sweeping the magnetic field from high fields to zero, an obvious drop of magnetoresistance can be observed at the beginning of the superconducting transition. This is inconsistent with our experimental results, so the flux trapping mechanism can be excluded.

### 3. Measurement-induced hysteresis

Possible artificial hysteresis in the measurement process has been considered. Due to the limited data acquisition rate of the computer, if the magnetic field sweep rate is faster than the acquisition rate, the collected data will lag and show hysteresis behavior. As shown in Fig. 3(b), the hysteresis in our measurement seems not sensitive to the field sweep rate, so this possibility can be ruled out.

### 4. Vortex physics

The vortex physics is one of the potential reasons for the magnetoresistance hysteresis behavior. Nevertheless, we want to point out that the vortex pinning effect is weak in our work. First of all, the *I-V* curves with opposite sweeping directions completely coincide with each other, without visible hysteresis behavior (Fig. 8), which is inconsistent with the vortex pinning model as reported in Ref. [52]. Moreover, the hysteresis behavior induced by the dynamic vortex is expected to be sensitive to the field sweeping rate. This is also inconsistent with our results in Fig. 3(b), where the hysteresis shows negligible dependence on the magnetic field sweeping rate. In addition, the absence of hysteresis behavior observed with no direct current applied [Fig. 2(a)] suggests the weak vortex pinning effect in our single-crystalline sample. Finally, with the large direct current applied, the hysteresis behaviors are observed in the finite resistance regime, where the vortices are already depinned by the large direct current. Therefore, the pinning strength of the vortex in our devices is very weak.

### 5. Transverse effects

Transverse effects have been considered in the measurement of $dV/dI$. Figure 3(b) presents the magnetoresistance hysteresis behavior measured in device S2, where the electrodes across the whole sample, as shown in Fig. 9, largely avoid the transverse signal represented by the transverse effects such as the Hall/Nerst effect. So, the transverse effects can be ruled out.

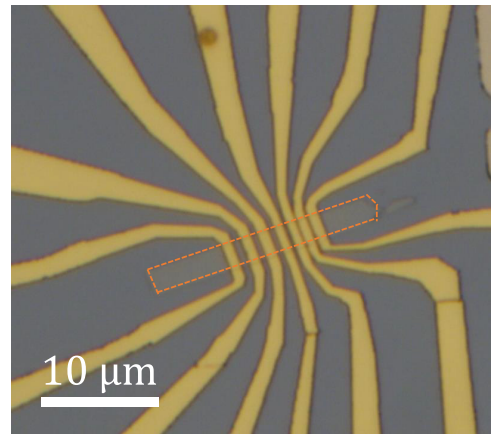

FIG. 9. Optical image of $CsV_3Sb_5$ device S2. The sample of device S2 is marked with dashed orange line, with thickness around 45 nm.